\documentclass[conference]{IEEEtran}
\IEEEoverridecommandlockouts

\usepackage{standalone}
\usepackage{tikz}
\usetikzlibrary{arrows.meta, patterns, decorations.pathmorphing}
\usepackage{cite}
\usepackage{amsmath,amssymb,amsfonts}
\usepackage{algorithmic}
\usepackage{graphicx}
\usepackage{textcomp}
\usepackage{xcolor}
\usepackage{siunitx}
\usepackage{booktabs,tabularx}
\usepackage{booktabs,threeparttable}
\usepackage{array}
\usepackage{ragged2e}

\def\BibTeX{{\rm B\kern-.05em{\sc i\kern-.025em b}\kern-.08em
		T\kern-.1667em\lower.7ex\hbox{E}\kern-.125emX}}
\begin{document}

\title{Analytically Benchmarked Spatial and Temporal Discretization Guidelines for Finite Element Simulation of Thermal Waves}

\author{
	\IEEEauthorblockN{
		Kaushal Arun Pareek\textsuperscript{1,*},
		Daniel May\textsuperscript{2},
		Peter Meszmer\textsuperscript{2,3},
		Bernhard Wunderle\textsuperscript{2,3,4,5},
		Mohamad Abo Ras\textsuperscript{1}
	}
	\IEEEauthorblockA{
		\textsuperscript{1}Berliner Nanotest and Design GmbH, Berlin, Germany\\
		\textsuperscript{2}Chair Materials and Reliability of Microsystems, Chemnitz University of Technology, Chemnitz, Germany\\
		\textsuperscript{3}Center for Materials, Architectures and Integration of Nanomembranes (MAIN), \\Chemnitz University of Technology, Chemnitz, Germany\\
		\textsuperscript{4}Micro Materials Center (MMC), Fraunhofer Institute for Electronic Nano Systems (ENAS), \\Chemnitz, Germany\\
		\textsuperscript{5}Center for Micro- and Nanotechnologies (ZfM), Chemnitz University of Technology, Chemnitz, Germany\\
		\textsuperscript{*}Corresponding author: pareek@nanotest.eu
	}
}

\IEEEpubid{\begin{minipage}{\textwidth}\ \\[12pt]\centering\footnotesize
		\copyright~2026 IEEE. Personal use of this material is permitted. Permission
		from IEEE must be obtained for all other uses, in any current or future media,
		including reprinting/republishing this material for advertising or promotional
		purposes, creating new collective works, for resale or redistribution to
		servers or lists, or reuse of any copyrighted component of this work in other
		works.
\end{minipage}}

\maketitle

\begin{abstract}
Finite element simulation is widely used to design and interpret thermal wave-based measurement systems. The accuracy of the results obtained from such simulations depends on how finely the model is resolved in space and time. The discretization resolution is often chosen case by case rather than following a well-defined guideline. This work recommends a discretization protocol for finite element simulation of thermal waves where the spatial discretization is scaled to the thermal diffusion length and the temporal discretization to the number of time steps per cycle. The associated discretization errors are quantified using three benchmark problems of increasing complexity. The spatial phase error converges at second order in the element size, reaching $\mathbf{0.14^\circ}$ at ten elements per diffusion length, whereas the temporal error decreases only at first order in the number of time steps per cycle; thus, the time step can be set from a target phase tolerance . Each error is assessed against a reference that isolates it. The benchmark case with subsurface defects shows the consequence of violating the recommended discretization protocol. 
\end{abstract}


\section{Introduction}
\label{sec:intro}
Periodic heating of a solid produces a diffusive thermal wave whose amplitude and phase can be utilized to characterize the material \cite{rosencwaigTheoryPhotoacousticEffect1976}. A one-dimensional analytical solution for the propagation of such a plane thermal wave into a semi-infinite solid in the $x$-direction is \cite{carslawConductionHeatSolids2011}, \cite{breitensteinLockinThermographyBasics2010}:
\begin{equation}
\label{eq:thermal_wave_eq}
	T(x, t) = A e^{-x/\mu} e^{j(\omega t - x/\mu)},
\end{equation}
with $\omega = 2\pi f$, where $f$ [\si{\hertz}] is the excitation frequency; $\mu = \sqrt{2\alpha/\omega}$ [\si{\meter}] is the thermal diffusion length with $\alpha$ being the thermal diffusivity [\si{\meter\squared\per\second}]; and $A$ is the amplitude of the oscillating temperature signal [\si{\kelvin}]. Thus, both the amplitude and phase of the thermal wave are influenced by $\mu$. Furthermore, the amplitude of the wave is strongly attenuated and decays by a factor of $e^{-2\pi}$ over one wavelength $\Lambda = 2\pi\mu$. This behavior is exploited in non-destructive techniques such as photoacoustic \cite{bennettThermalWaveInterferometry1982} and lock-in thermography (LIT) \cite{wuLockinThermographyNondestructive1998}. In each case, a periodic excitation is applied to the surface, and the resulting oscillating temperature field is recorded. Of the two observables, the phase is generally preferred for quantitative evaluation as it is largely independent of local variations in surface emissivity and of inhomogeneous illumination \cite{busseThermalWaveImaging1992}, whereas amplitude is affected by both. From~\eqref{eq:thermal_wave_eq}, the phase at a depth $x$ is $\varphi = -x/\mu$; thus, the depth is measured in units of the thermal diffusion length. Since the excitation frequency sets $\mu$, the physical length scale is a property of the measurement setup. An excitation frequency sweep, which is the basis of depth-resolved evaluation, therefore traverses a corresponding wide range of length scales. Any numerical treatment of such a problem has to account for this frequency-dependent scale in its discretization. 

\IEEEpubidadjcol
Equation~\eqref{eq:thermal_wave_eq} is an exact solution for a semi-infinite solid. In a sample of finite thickness, the wave reflects at the rear boundary and the surface response follows from a geometric series over multiple internal reflections \cite{bennettThermalWaveInterferometry1982, gleiterThermalWaveInterference2010}. Both solutions are complex-valued and therefore provide two independent measures against which a numerical solution can be assessed; of these, the phase is the more demanding, as it is not normalized by the signal magnitude and stays sensitive where the amplitude has decayed. Beyond one-dimensional layered geometry, closed-form solutions are no longer available for defects of finite depth, where heat flows around the feature, and the response is three-dimensional. In this regime, quantitative depth estimation relies on relations established numerically or empirically, such as the blind frequency and phase contrast methods \cite{delanthabettuDefectDepthQuantification2015, baiEvaluationDefectsComposite2001}, and these relations are only conditionally valid: the observed phase contrast depends on the lateral extent of the defect as well as its depth \cite{wallbrinkEffectSizeQuantitative2007, maQuantitativeDepthEstimation2025}. Where such relations are derived from simulation, the numerical model itself becomes the reference against which depth estimates are calibrated. The closed-form solutions of the one-dimensional case are widely used to interpret measurements and to invert them for depth, but rarely as references against which the discretization of the numerical models that replace them is verified.

Finite element (FE) simulation is the established route to configurations for which no closed form exists, and the thermal diffusion length is routinely used to make such models general. Dimensionless formulation expresses defect depth, crack length, and probe position in units of $\mu$, rendering the analysis independent of the material properties and excitation frequency \cite{sagarduy-marcosDimensionlessNumericalSensitivity2024, rodriguez-aseguinolazaQuantifyingWidthAngle2021}. The discretization of these same models is not tied to that scale. Chatterjee and Tuli \cite{chatterjeePredictionBlindFrequency2013} presented a study for blind frequency prediction using numerical simulations based on electro-thermal models where the recommended element length was a fraction of $\mu$, and exponential slicing was recommended for thick bodies. However, they did so for a one-dimensional lumped electro-thermal network, without any convergence study or comparison against a closed-form solution. Studies employing a range of excitation frequencies or harmonic-rich excitation signals generally apply a single element size and solution time step across all frequencies, resulting in coarse resolving of $\mu$ at high excitation frequencies. Furthermore, convergence w.r.t to mesh and solution time is typically reported as an implementation detail, without reference to the excitation frequency or the diffusion length \cite{maQuantitativeDepthEstimation2025, riazNovelApproachSubsurface2026, schmidtThermalSimulationDefect2009, pengModellingLockinThermography2013}.

General-purpose FE resources recommend $\Delta t = \Delta x ^2/4\alpha$ for transient thermal analysis \cite[Section~3.4.3.2]{ansysThermalAnalysisGuide2025}, a relationship between the largest element size $\Delta x$ and the smallest time step $\Delta t$. This bound is derived from the discrete maximum principle rather than from accuracy. Rank et al. \cite{rankImportanceDiscreteMaximum1983} showed that in FE approximation, small timesteps can produce physically unreasonable oscillatory results, contrary to the general expectation that reducing the step size only improves the accuracy. The factors governing the solution step size for the diffusion problem were subsequently analyzed by Thomas et al. \cite{thomasAnalysisFactorsThat1998}. Satisfying these criteria prevents spurious oscillations but says nothing about whether the temporal and spatial discretization is sufficient to represent a thermal wave. Furthermore, the bounds reference neither the excitation frequency nor the diffusion length. Thus, providing no indication as to how the discretization should be adapted as the excitation frequency is swept, as is the case in LIT. Nor do they address harmonic analysis, where no time integration is required. 

The missing element is therefore not a new numerical method but a criterion referred to the scale that governs the physics. This work establishes an analytically benchmarked discretization protocol for FE simulations of thermal waves and verifies it against closed-form solutions at three levels of complexity: a one-dimensional semi-infinite solid, samples ranging from thermally thin to thermally thick with internal reflections, and a three-dimensional model containing a subsurface defect. The contributions of this paper are as follows:
\begin{enumerate}
	\item The phase error associated with the diffusion-length-scaled element size $\Delta x = \mu/a$ is quantified as a function of the scaling factor $a$ and benchmarked against closed-form solutions.
	\item An accuracy-driven bound on solution time step $\Delta t$ is established for transient thermal analysis. The spatial and temporal errors are shown to be separable, requiring the temporal error to be referenced against the harmonic solution to avoid apparent convergence through error cancellation. 
	\item The conditions under which harmonic analysis is sufficient and transient analysis is required, are identified.
	\item Protocol violation is shown to distort the relationship between defect depth and blind frequency in a three-dimensional model.
\end{enumerate}
The scope of this work is verification against analytical solutions; no experimental comparisons are presented. The transferability of the protocol rests on the thermal diffusion length of the governing equations.  

\section{Analytical Benchmark Problems}
\label{sec:benchmark}
Three benchmark problems of increasing physical complexity are used to quantify the FE model's discretization error. For each problem, this section states the closed-form solution and the assumptions the FE model must reproduce for meaningful comparison. Section~\ref{sec:results} reports the error metric. Table~\ref{tab:benchmarks} summarizes the three problems and the corresponding schematics are shown in Fig.\ref{fig:benchmarks}.

\begin{table*}[t]
	\centering
	\caption{Analytical benchmark problems used to quantify the finite element discretisation error. Each closed-form solution defines the verification target for the corresponding results subsection}
	\label{tab:benchmarks}
	\renewcommand{\arraystretch}{1.3}
	\begin{tabularx}{\textwidth}{@{}
			>{\raggedright\arraybackslash}p{2.4cm}
			>{\raggedright\arraybackslash}p{3.0cm}
			>{\raggedright\arraybackslash}X
			>{\raggedright\arraybackslash}p{3.4cm}
			c @{}}
		\toprule
		\textbf{Benchmark} & \textbf{Closed-form solution} & \textbf{Assumptions the FE model must reproduce} & \textbf{Verification metric} & \textbf{Results} \\
		\midrule
		Semi-infinite solid (Section \ref{subsec:semi_infinite_solid}) & Equations~\eqref{eq:thermal_wave_eq} and \eqref{eq:surf_response}, periodic surface heat flux & Ideal flux boundary condition; truncation length $L_{\infty} \geq 6\mu$; temperature-independent properties
		& $\varphi(0)=-\pi/4$;\; $\dfrac{\mathrm{d}\varphi}{\mathrm{d}x}=\dfrac{\mathrm{d}\ln|A|}{\mathrm{d}x}=-\dfrac{1}{\mu}$; spatial and temporal convergence orders
		& Section \ref{subsec:res_semi_infinite_solid} \\
		\addlinespace
		Finite thickness (Section \ref{subsec:internal_reflection}) & Equations~\eqref{eq:bp_surface_phase} and \eqref{eq:bp_surface_phase_simplified} & Adiabatic rear face ($R_b=1$); front flux condition identical to Section \ref{subsec:semi_infinite_solid}; one-dimensional
		& Phase MAE against~\eqref{eq:bp_surface_phase} & Section \ref{subsec:res_internal_reflection} \\
		\addlinespace
		Subsurface defect (Section \ref{subsec:blind_freq}) & 1-D limit $d=\tfrac{\pi}{2}\mu(f_b)$, Equation~\eqref{eq:blind_freq}; none for finite lateral extent & One-dimensional reflection; finite lateral extent ($r/\mu \sim 1$) and finite sound-region reference (no closed form)
		& Mesh-strategy dependence of the recovered $d$--$\mu(f_b)$ relation; departure from the 1-D identity ($C<\pi/2$) & Section \ref{subsec:res_blind_freq} \\
		\bottomrule
	\end{tabularx}
\end{table*}
\subsection{Semi-Infinite Solid}
\label{subsec:semi_infinite_solid}
The first benchmark isolates the discretization error in the absence of any surface defect. Applying a periodic surface heat flux $q_0 e^{j\omega t}$ at the surface of the semi-infinite solid governed by~\eqref{eq:thermal_wave_eq} gives the surface temperature response \cite{carslawConductionHeatSolids2011, breitensteinLockinThermographyBasics2010} 
\begin{equation}
	\label{eq:surf_response}
	T(0, t) = \frac{q_0}{e\sqrt{\omega}}e^{j(\omega t - \pi/4)}, \quad e = \sqrt{k\rho c_p},
\end{equation}
where $q_0$ is the heat flux [\si{\watt\per\meter\squared}], $k$ is the thermal conductivity [\si{\watt\per\meter\per\kelvin}], $\rho$ is the density [\si{\kilogram\per\cubic\metre}], $c_p$ is the specific heat [\si{\joule\per\kilogram\per\kelvin}], and $e$ is the thermal effusivity [\si{J.m^{-2}.K^{-1}.s^{-1/2}}]. The phase shift between the modulated power and the resulting surface temperature modulation is $-\pi/4$, which emerges from the solution and therefore constitutes a scalar test for the discretization study. In addition, from~\eqref{eq:thermal_wave_eq} both the phase and the logarithmic amplitude decreases linearly with depth, $\mathrm{d}\varphi/\mathrm{d}x = \mathrm{d}\ln|A|/\mathrm{d}x = -1/\mu$, providing two further verification quantities.
\subsection{Finite Thickness: Internal Reflections}
\label{subsec:internal_reflection}
Real samples can seldom be modeled as a semi-infinite solid; thus, the FE domain is necessarily of finite length $L$. The second benchmark introduces a single reflecting boundary and, with it, thermal-wave interference \cite{gleiterThermalWaveInterference2010}. For a slab of finite thickness $L$ whose front and rear faces reflect with coefficients $R_g$ and $R_b$, where $R = \frac{e_s-e'}{e_s+e'}$ with $e_s$ being the effusivity of the sample and $e'$ the effusivity of the medium beyond each face, the forward and backward waves sum to
\begin{equation}
	\label{eq:bp_general}
	T(x, t) = \frac{q_0}{k\sigma}\,
	\frac{e^{-\sigma x}+R_b\, e^{-\sigma (2L-x)}}{1-R_gR_b\, e^{-2\sigma L}}\;e^{j\omega t},
\end{equation}
where $\sigma = \frac{1+j}{\mu}$. Equation~\eqref{eq:bp_general} is the closed form of the Bennett-Patty multiple-reflection series \cite{bennettThermalWaveInterferometry1982}, whose $n\rightarrow\infty$ sum over successive round trips converges for every excitation frequency because each traversal attenuates the wave by $|R_gR_be^{-2\sigma L}| < 1$. Evaluating~\eqref{eq:bp_general} at the observed surface of $x = 0$ gives 
\begin{equation}
	\label{eq:bp_general_surface}
	T(0,t) = \frac{q_0}{k\sigma}\frac{1+R_b e^{-2\sigma L}}{1-R_gR_b e^{-2\sigma L}}e^{j\omega t}.
\end{equation}
Setting $R_b=0$ describes a rear interface matched to the sample, i.e., no wave returns, and~\eqref{eq:bp_general_surface} collapses to the semi-infinite response of~\eqref{eq:surf_response}. For a sample in air, $R_g=R_b=1$, gives
\begin{equation}
	\label{eq:bp_coth}
	T(0,t) = \frac{q_0}{e\sqrt{\omega}}e^{-j\pi/4}\coth(\sigma L)e^{j\omega t}.
\end{equation}
Thus, the resulting surface phase separates additively into a contribution from the source and one from the interference
\begin{equation}
	\label{eq:bp_surface_phase}
	\varphi(0) = -\pi/4 + \arg \coth(\sigma L).
\end{equation}
The interference term can be further simplified as 
\begin{equation}
	\label{eq:bp_surface_phase_simplified}
	\arg \coth(\sigma L) = -\arctan\left(\frac{\sin u}{\sinh u}\right), \quad u = \frac{2L}{\mu}.
\end{equation}
The limiting values of~\eqref{eq:bp_surface_phase_simplified} and resulting surface phase values are listed in Table \ref{tab:phasevalues}. For the thermally thick case ($L\gg\mu$), the interference term vanishes and the semi-infinite phase of $-\pi/4$ is recovered; in the thermally thin case ($L\ll\mu$), the surface phase approaches $-\pi/2$. The verification metric at this level is the mean absolute error (MAE) of the simulated phase values against~\eqref{eq:bp_surface_phase} over the range $L/\mu=0.3-7.2$.

\begin{table}[!t]
	\centering
	\begin{threeparttable}
		\caption{Limiting values of the interference phase $\arg\coth(\sigma L)$ and the resulting surface phase $\varphi(0)=-\pi/4+\arg\coth(\sigma L)$, with $u=2L/\mu$}
		\label{tab:phasevalues}
		\renewcommand{\arraystretch}{1.3}
		\begin{tabular}{@{}l c c c@{}}
			\toprule
			\textbf{Regime} & \textbf{Condition} &
			$\arg\coth(\sigma L)$ & $\varphi(0)$ \\
			\midrule
			Thermally thick  & $u\to\infty$ & $0$   & $-\pi/4$ \\
			Thermally thin   & $u\to 0$     & $-\pi/4$ & $-\pi/2$ \\
			Blind frequency  & $u=\pi$      & $0$   & $-\pi/4$ \\
			\bottomrule
		\end{tabular}
		\begin{tablenotes}[flushleft]\footnotesize
			\item The thermally thick limit and the blind frequency both yield $\varphi(0)=-\pi/4$: the former is an asymptote, the latter a zero crossing above which the phase contrast inverts sign.
		\end{tablenotes}
	\end{threeparttable}
\end{table}

\subsection{Subsurface Defect: The Blind-Frequency Relation}
\label{subsec:blind_freq}
The third benchmark applies the interference model of Section \ref{subsec:internal_reflection} to a subsurface defect at depth $d$. In LIT, the defect is detected through phase contrast between the defect region and a sound region, which, since the same flux drives both, removes the common $-\pi/4$:

\begin{equation}
	\label{eq:phase_contrast}
	\Delta \varphi = \arg \coth(\sigma d) - \arg \coth(\sigma L),
\end{equation}
where $L$ is the thickness of the sound region. The defect becomes invisible at the frequency where the two phases coincide, $\Delta \varphi = 0$. When the sound region is thermally thick, $\arg \coth(\sigma L) \rightarrow 0$, and the blind-frequency condition reduces to a zero of the defect term alone. From~\eqref{eq:bp_surface_phase_simplified}, $\arg \coth(\sigma d) = -\arctan\left(\sin u/ \sinh u\right)$ with $u = 2d/\mu$, this occurs when $\sin(2d/\mu) = 0$. The first nonzero root, $2d/\mu = \pi$, defines the fundamental blind frequency $f_b$, at which

\begin{equation}
	\label{eq:blind_freq}
	d = \frac{\pi}{2}\mu(f_b), \quad \mu(f_b) = \sqrt{\alpha/\pi f_b}.
\end{equation}
The constant $C = \pi/2$ relating the defect depth to the diffusion length at blind frequency is thus an analytical identity of the one-dimensional model. When the sound region is of finite thickness, its own interference term does not vanish, and the blind-frequency condition shifts to a geometry-dependent value. This artifact in reference region selection (thermally thick sound region at $f_b$ or sound region of finite thickness $L$) contributes to the depth-dependent constant observed experimentally \cite{delanthabettuDefectDepthQuantification2015}, with finite lateral extent of the defect contributing a bias \cite{maQuantitativeDepthEstimation2025}. Equation~\eqref{eq:phase_contrast} further assumes, through the one-dimensional model, a defect of lateral extent much greater than $\mu$; for the finite flat-bottom holes modeled here, heat conducts laterally around the defect, and no closed-form solution exists \cite{wallbrinkEffectSizeQuantitative2007}. The three-dimensional flat-bottom-hole defect is therefore not treated as an analytical benchmark but as a demonstration that the analytically validated discretization protocol extends to a geometry without a closed form.

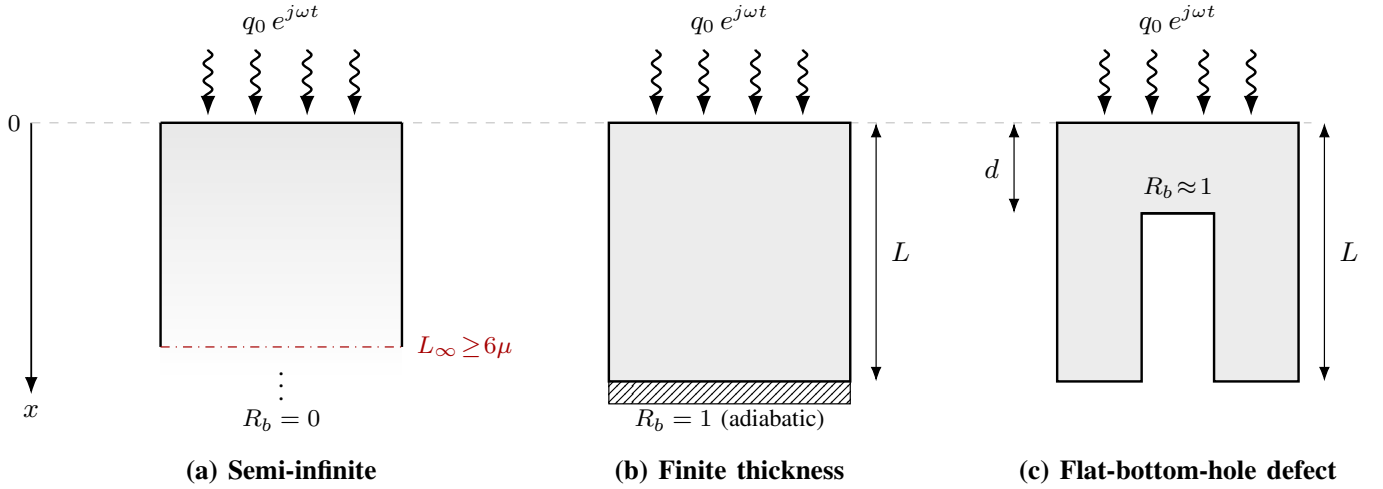
\begin{figure*}[t]
	\centering
	\resizebox{\textwidth}{!}{\begin{tikzpicture}[
  >=Latex,
  flux/.style={->, thick, decorate,
    decoration={snake, amplitude=0.5mm, segment length=2mm, post length=1.3mm}},
  dimline/.style={<->, >=Latex, thin},
  axis/.style={->, semithick},
  every node/.style={font=\small}
]

\def\w{2.8}      
\def\H{3.0}      
\def\dx{5.2}     

\draw[axis] (-1.5,0) -- (-1.5,-\H-0.15) node[below] {$x$};
\node[left, font=\footnotesize] at (-1.5,0) {$0$};
\draw[gray!45, dashed, thin] (-1.5,0) -- (2*\dx+\w+0.15,0);

\begin{scope}[shift={(0,0)}]
  \shade[top color=gray!18, bottom color=white] (0,0) rectangle (\w,-\H);
  \draw[thick] (0,0) -- (\w,0);                 
  \draw[thick] (0,0) -- (0,-\H+0.4);
  \draw[thick] (\w,0) -- (\w,-\H+0.4);
  \foreach \x in {0.55,1.1,1.7,2.25}{\draw[flux] (\x,0.85) -- (\x,0.08);}
  \node at (\w/2,1.2) {$q_0\,e^{j\omega t}$};
  \draw[dash dot, red!65!black] (0,-\H+0.4) -- (\w,-\H+0.4);
  \node[red!65!black, right, font=\footnotesize] at (\w+0.05,-\H+0.4)
        {$L_\infty\!\ge\!6\mu$};
  \node at (\w/2,-\H+0.05) {$\vdots$};
  \node[font=\footnotesize] at (\w/2,-\H-0.45) {$R_b = 0$};
  \node at (\w/2,-\H-1.05) {\textbf{(a) Semi-infinite}};
\end{scope}

\begin{scope}[shift={(\dx,0)}]
  \fill[gray!15] (0,0) rectangle (\w,-\H);
  \draw[thick] (0,0) rectangle (\w,-\H);
  \draw[pattern=north east lines, pattern color=black]
        (0,-\H) rectangle (\w,-\H-0.26);
  \node[font=\footnotesize] at (\w/2,-\H-0.45) {$R_b = 1$ (adiabatic)};
  \foreach \x in {0.55,1.1,1.7,2.25}{\draw[flux] (\x,0.85) -- (\x,0.08);}
  \node at (\w/2,1.2) {$q_0\,e^{j\omega t}$};
  \draw[dimline] (\w+0.3,0) -- (\w+0.3,-\H);
  \node[right] at (\w+0.35,-\H/2) {$L$};
  \node at (\w/2,-\H-1.05) {\textbf{(b) Finite thickness}};
\end{scope}

\begin{scope}[shift={(2*\dx,0)}]
  \def\d{1.05}     
  \def\hw{0.42}    
  \fill[gray!15] (0,0) rectangle (\w,-\H);
  \fill[white] ({\w/2-\hw},-\H) rectangle ({\w/2+\hw},-\d);
  \draw[thick]
      (0,0) -- (\w,0) -- (\w,-\H)
      -- ({\w/2+\hw},-\H) -- ({\w/2+\hw},-\d)
      -- ({\w/2-\hw},-\d) -- ({\w/2-\hw},-\H)
      -- (0,-\H) -- cycle;
  \draw[thick] ({\w/2-\hw},-\d) -- ({\w/2+\hw},-\d);
  \node[font=\footnotesize] at (\w/2,-\d+0.30) {$R_b\!\approx\!1$};
  \foreach \x in {0.55,1.1,1.7,2.25}{\draw[flux] (\x,0.85) -- (\x,0.08);}
  \node at (\w/2,1.2) {$q_0\,e^{j\omega t}$};
  \draw[dimline] (-0.5,0) -- (-0.5,-\d);
  \node[left] at (-0.55,-\d/2) {$d$};
  \draw[dimline] (\w+0.3,0) -- (\w+0.3,-\H);
  \node[right] at (\w+0.35,-\H/2) {$L$};
  \node at (\w/2,-\H-1.05) {\textbf{(c) Flat-bottom-hole defect}};
\end{scope}

\end{tikzpicture}}
	\caption{The three analytical benchmark problems, driven by a common surface heat flux $q_0e^{j\omega t}$ and observed through the surface phase at $x=0$. (a) Semi-infinite solid: the reflectionless limit ($R_b=0$), realized in the finite element model by a truncation length $L_\infty \geq 6\mu$ that suppresses the boundary reflection. (b) Finite slab of thickness $L$ with an adiabatic rear face ($R_b=1$), producing thermal-wave interference. (c) Flat-bottom-hole defect of depth $d$ in a sample of thickness $L$, the hole drilled from the rear face}
	\label{fig:benchmarks}
\end{figure*}

\subsection{Error Metrics and Conventions}
\label{subsec:conventions}
MAE is reported for the discretization error between simulated amplitude and phase values against the analytical equations presented in this section. Phase values are in degrees and are measured relative to the excitation, with a negative value representing a phase lag. Phase contrast between two regions is reported as $\varphi_{\mathrm{defect}}-\varphi_{\mathrm{sound}}$ consistent with~\eqref{eq:phase_contrast}.


\section{Finite Element Methodology}
\label{sec:fe_methodology}
This section explains how the FE models implement the benchmark conditions in Section~\ref{sec:benchmark}, the two FE analyses compared in Section~\ref{sec:results}, and the spatial and temporal discretization protocol. 

\subsection{Model Configuration and Boundary Conditions}
\label{subsec:model_config}
The semi-infinite (Section~\ref{subsec:semi_infinite_solid}) and finite-slab (Section~\ref{subsec:internal_reflection}) benchmarks are modeled by a single-element-cross-section column of linear thermal-conducting brick elements. The side faces are thermally insulated; thus, conduction occurs along the depth. A two-dimensional axisymmetric model consisting of a coaxial flat-bottom hole is used for the subsurface-defect study (Section~\ref{subsec:blind_freq}). All the models use ANSYS thermal elements. The excitation is a periodic surface heat flux of frequency $f$ applied uniformly on the top surface with adiabatic conditions on the rear surface. Material properties of the used material are listed in Table~\ref{tab:materials}.
\begin{table}[!t]
	\renewcommand{\arraystretch}{1.2}
	\caption{Thermal properties of the benchmark materials}
	\label{tab:materials}
	\centering
	\footnotesize
	\setlength{\tabcolsep}{4pt}
	\begin{tabular}{l c c c c}
		\toprule
		\textbf{Material} & $k$ & $c_p$ & $\rho$ & $\alpha$ \\
		& [$\mathrm{W\,m^{-1}K^{-1}}$] & [$\mathrm{J\,kg^{-1}K^{-1}}$]
		& [$\mathrm{kg\,m^{-3}}$] & [$\mathrm{m^{2}\,s^{-1}}$] \\
		\midrule
		AlMg3 & 132 & 897 & 2680 & 5.49$\times10^{-5}$ \\
		Al    & 237 & 903 & 2702 & 9.71$\times10^{-5}$ \\
		\bottomrule
		\multicolumn{5}{@{}p{0.98\columnwidth}@{}}{
			$\alpha$ is derived as $k/(\rho c_p)$}
	\end{tabular}
\end{table}

\subsection{Analysis Type}
\label{subsec:analysis_type}
Two analyses are compared: the harmonic and transient thermal analyses. The harmonic analyses (frequency-domain) directly return the complex steady-state field at $f$, with amplitude and phase given by its modulus and argument. As harmonic analysis doesn’t require any time integration, it isolates the spatial discretization error and is the reference against which the transient thermal analysis is evaluated. It is used for all frequency-sweep results. The transient analysis integrates $\mathbf{C}\dot{\mathbf{T}} + \mathbf{K}\mathbf{T} = \mathbf{F}$ by implicit generalised-trapezoidal ($\theta$) rule, for which ANSYS defaults to $\theta = 1$ (backward Euler method)\cite[Section~15.2]{ansysTheoryRef2025}. Transient analysis is used for the temporal convergence study and requires a warm-up of a few excitation periods before it reaches the periodic steady state. Table~\ref{tab:config} lists the per-benchmark model configuration.

\begin{table*}[!t]
	\renewcommand{\arraystretch}{1.2}
	\setlength{\tabcolsep}{4pt}
	\caption{Finite element model configuration for the three benchmark problems. In all cases the excitation is a periodic surface heat flux; the Excitation column lists the frequency treatment}
	\label{tab:config}
	\centering
	\small
	\begin{tabular}{@{}
			>{\RaggedRight\arraybackslash}p{0.135\linewidth}
			>{\RaggedRight\arraybackslash}p{0.375\linewidth}
			>{\RaggedRight\arraybackslash}p{0.075\linewidth}
			>{\RaggedRight\arraybackslash}p{0.205\linewidth}
			>{\RaggedRight\arraybackslash}p{0.125\linewidth}
			@{}}
		\toprule
		\textbf{Benchmark} & \textbf{Geometry} & \textbf{Material} & \textbf{Analysis type} & \textbf{Excitation} \\
		\midrule
		Semi-infinite (Section~\ref{subsec:semi_infinite_solid}) &
		1-D column, $L_{\infty}=6\mu$, insulated sides, adiabatic far face &
		AlMg3 &
		Harmonic (spatial); transient (temporal) &
		Single frequency \\
		\addlinespace
		Finite slab (Section~\ref{subsec:internal_reflection}) &
		1-D column, $L=2\,\mathrm{mm}$, insulated sides, adiabatic rear ($R=1$) &
		AlMg3 &
		Harmonic &
		Frequency swept \\
		\addlinespace
		Flat-bottom hole (Section~\ref{subsec:blind_freq}) &
		2-D axisymmetric, sample radius $2\,\mathrm{mm}$, $L=6\,\mathrm{mm}$;
		coaxial hole radius $0.5\,\mathrm{mm}$, depth $d$ swept &
		Al &
		Harmonic &
		Swept per depth \\
		\bottomrule
	\end{tabular}
\end{table*}

\subsection{Spatial and Temporal Discretization}
\label{subsec:discretization_protocol}
Element size follows $\Delta x = \mu/a$, with $a$ being the number of elements per diffusion length. A mapped uniform meshing is adopted for all the simulations of all the benchmark problems. For the frequency-swept cases, the mesh is regenerated at each $f$ so that $a$ remains fixed as $\mu$ changes. In the subsurface-defect case (Section~\ref{subsec:blind_freq}), the $\mu$-scaled mesh is compared against a mesh of constant element size, to evaluate the consequence of resolving the field by geometry rather than the diffusion length.

The study isolates discretization errors by separating the spatial and temporal discretization channels. Spatial resolution is studied with harmonic analysis, which carries no temporal error, by varying $a$ at a fixed frequency. Temporal resolution is then studied with transient thermal analysis at a spatial resolution fine enough that spatial error is negligible, by fixing $a$ and varying the number of samples per period, $N_t$. Transient thermal simulations use the ANSYS default backward-Euler scheme ($\theta = 1$), and automatic time-stepping is disabled so the prescribed time step is imposed exactly. As implicit integration is unconditionally stable, the step size is chosen for accuracy rather than stability. The adequate $a$ and $N_t$ are established in Section~\ref{sec:results}.

\subsection{Amplitude and Phase Extraction}
\label{subsec:amp_&_phase}
Amplitude and phase values are extracted per node. For harmonic analysis, amplitude is the modulus of the complex field, and phase is its argument relative to the excitation. For transient analysis, the initial warm-up (5 to 7 periods) is discarded, and the amplitude and phase are obtained from the discrete Fourier transform of the settled nodal signal at the excitation frequency.


\section{Results and Discussions}
\label{sec:results}
The discretization protocol described in Section \ref{sec:fe_methodology} is verified against the three analytical benchmarks of Table~\ref{tab:benchmarks} in order of increasing complexity. Section~\ref{subsec:res_semi_infinite_solid} isolates the spatial and temporal discretization errors independently and helps establish the mesh and time-stepping criteria. Section~\ref{subsec:res_internal_reflection} confirms that the proposed spatial discretization guideline reproduces the frequency-dependent interference in a finite-thickness solid. Lastly, Section~\ref{subsec:res_blind_freq} demonstrates the protocol's applicability to a different material with subsurface defects in a three-dimensional geometry for which no closed-form solution exists, showing that violating the protocol alters the practical depth measurement.

\subsection{Semi-Infinite Solid}
\label{subsec:res_semi_infinite_solid}
\begin{figure}
	\centering
	\includegraphics[width=\columnwidth]{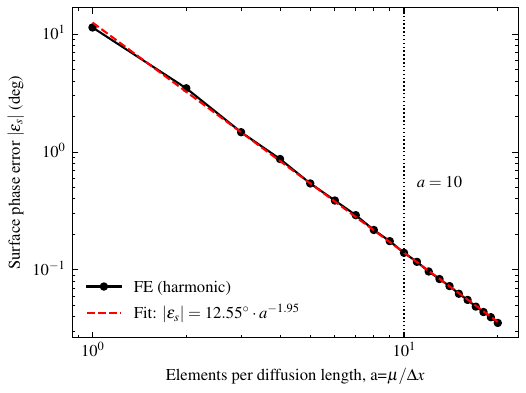}
	\caption{Spatial convergence of the harmonic surface phase against the semi-infinite analytical benchmark. The surface phase error decreases at approximately second order, giving $|\varepsilon_s|\approx 0.14^\circ$ at the adopted resolution of $a=10$}
	\label{fig:spatial_conv}
\end{figure}

\noindent\textit{Spatial convergence}:
The harmonic simulation results were evaluated for a mapped mesh of $a = \mu/\Delta x$ elements per diffusion length, with $a$ swept from 1 to 20 at a fixed excitation frequency of 20 Hz. The surface phase error relative to the analytical value $\varphi(0) = -45^\circ$~\eqref{eq:surf_response} is show in Fig.~\ref{fig:spatial_conv}. The spatial surface phase error ($\varepsilon_s$) decreases at approximately second order (fitted exponent 1.95). In contrast, the amplitude error (not shown) converged to the analytical value on a relatively coarse mesh ($a = 5$). An example of this asymmetric behavior is illustrated at $a = 5$, where the amplitude already agrees with the analytical value while the phase still deviates by $0.54^\circ$. Thus, any spatial convergence value must be phase error-based, as amplitude error alone leads to false spatial convergence. For $a \geq 10$, the spatial phase error is below $0.15^\circ$, which is well below the phase noise of typical measurements and is adopted for the subsequent simulations.

\begin{figure}
	\centering
	\includegraphics[width=\columnwidth]{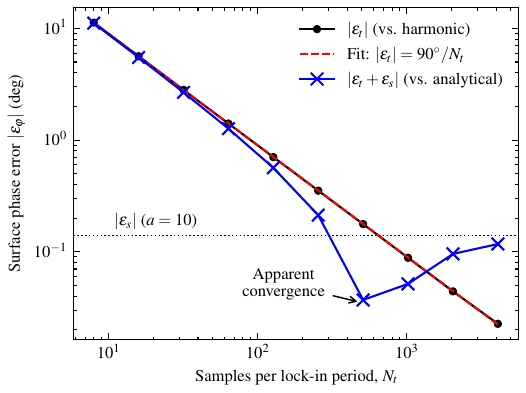}
	\caption{Temporal convergence of the transient surface phase at fixed mesh ($a=10$). Referenced to the harmonic solution at the same mesh, the temporal error follows first-order convergence $|\varepsilon_t|=90^\circ/N_t$. Referenced to the analytical solution, the error shows an apparent convergence near $N_t\approx 500$ where the temporal and spatial errors cancel before rising to the spatial error floor}
	\label{fig:temporal_conv}
\end{figure}

\noindent\textit{Temporal convergence}:
Thermal transient simulations were evaluated at a fixed mesh size with $a = 10$ elements per diffusion length and an excitation frequency of 20 Hz. The number of samples per period $N_t$ was parameterized to give varying solution time-stepping $\Delta t = 1/(f\times N_t)$ for the transient simulation. The values of $N_t$ were swept from 8 to 4096 samples per period. The transient solution surface error contains both the spatial error ($\varepsilon_s$) already characterized and a temporal contribution ($\varepsilon_t$).
\begin{align}
	\varepsilon_s &= \varphi_{\text{harmonic}}(0) - \varphi(0), \label{eq:spatial_error} \\
	\varepsilon_t &= \varphi_{\text{transient}}(0) - \varphi_{\text{harmonic}}(0), \label{eq:temporal_error} \\
	\varepsilon_s + \varepsilon_t &= \varphi_{\text{transient}}(0) - \varphi(0), \label{eq:total_error}
\end{align}
where $\varphi_{\text{harmonic}}(0)$ is the surface phase value from the harmonic analysis and $\varphi_{\text{transient}}(0)$ is the surface phase value from transient analysis. As seen from~\eqref{eq:temporal_error}, referencing to harmonic isolates $\varepsilon_t$; referencing to analytical~\eqref{eq:total_error} retains both the errors. Fig.~\ref{fig:temporal_conv} shows the error in transient analysis. Referenced to the harmonic analysis, the temporal phase error follows a first-order law, $\varepsilon_t = 90^\circ/N_t$. This first-order convergence is consistent with the backward-Euler integrator's first-order accuracy. The same fit was recovered at $a=8$, thus confirming that the temporal error is independent of the mesh. When referencing the transient solution to the analytical solution, an apparent convergence is observed at $N_t \approx 500$. This is because the spatial error $\varepsilon_s$ is negative while the temporal error $\varepsilon_t$ is positive; the opposite signs cause the two contributions to cancel at a particular $N_t$, producing the apparent convergence as seen in Fig.~\ref{fig:temporal_conv}. Furthermore, an increasing $N_t$ increases the error, which converges toward $ 0.14^\circ$, the spatial error in Fig.~\ref{fig:spatial_conv}. Thus, refining the solution time step beyond this point moves the solution away from the analytical value and toward the harmonic solution on the same mesh, which retains the spatial error. 
The transient solution therefore cannot reach the analytical phase by temporal refinement alone; for a given mesh, the best a transient solution can attain is the harmonic solution at the same mesh. Reaching the analytical value requires both spatial and temporal refinement at the same time. Furthermore, as shown in Fig.~\ref{fig:temporal_conv}, validating a transient result against an analytical solution is unreliable, because the simulation can appear converged through error cancellation. Additionally, it was noted that the bound of~\cite{rankImportanceDiscreteMaximum1983}, which for the presented scheme corresponds to $N_t\le6\pi a^2$, was exceeded at the finest time steps and no unreasonable oscillatory results were observed. 

\begin{figure}
	\centering
	\includegraphics[width=\columnwidth]{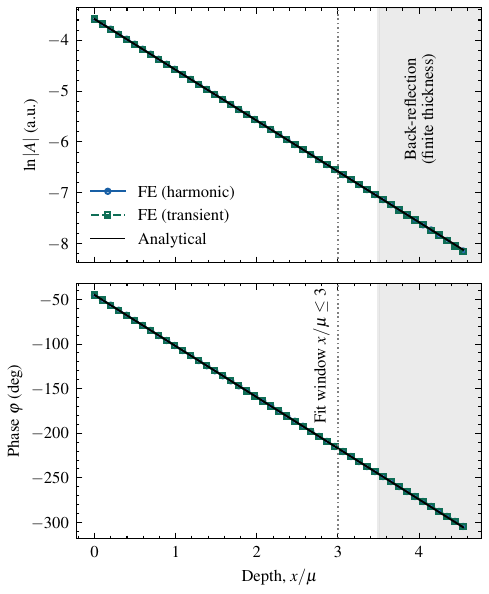}
	\caption{Log-amplitude (top) and unwrapped phase (bottom) along the propagation direction for the harmonic ($a = 10$) and transient ($a = 10, N_t = 4096$) solution against the semi-infinite analytical benchmark. Both quantities decay linearly at $-1/\mu$. The shaded region represents the onset of finite-thickness back-reflection}
	\label{fig:depth_profile}
\end{figure}

\noindent\textit{Slope verification}:
Fig.~\ref{fig:depth_profile} shows the amplitude and phase values along the propagation direction. Both harmonic ($a=10$) and transient ($a=10, N_t = 4096$) match the analytical decay of~\eqref{eq:thermal_wave_eq}. The log-amplitude and phase profiles fall linearly with a slope of $-1/\mu$. Fitted over the reflection-free window $x/\mu\le 3$, both simulations agree with the analytical slope and are indistinguishable. Beyond $x/\mu\approx 3.5$, both profiles depart from the analytical curve as the reflected wave from the insulated rear end contaminates the response.

\subsection{Finite-Thickness Slab}
\label{subsec:res_internal_reflection}
\begin{figure}
	\centering
	\includegraphics[width=\columnwidth]{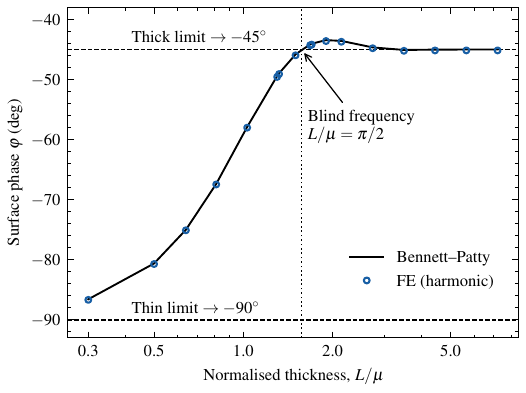}
	\caption{Surface phase of the finite-thickness slab against Bennett-Patty solution~\eqref{eq:bp_surface_phase}. The normalized thickness $L/\mu$ is varied from 0.3 to 7.2 by varying the excitation frequency $f$ and thus the thermal diffusion length $\mu$ at fixed mesh $a=10$. The FE solution reproduces the analytical response with a phase MAE of $0.14^\circ$, including the values of Table~\ref{tab:phasevalues}}
	\label{fig:bennett_patty}
\end{figure}

The finite-thickness slab changes the length scale from the previous section and tests whether the meshing strategy reproduces the frequency-dependent interference of a finite-thickness geometry. A single sample of thickness \qty{2}{\milli\meter} was modeled with boundary conditions as described in Table~\ref{tab:config}. The excitation frequency was swept so that $L/\mu$ varies from 0.3 to 7.2. The mesh was rescaled at each frequency to hold $a=10$. Fig.~\ref{fig:bennett_patty} compares the harmonic surface phase to the analytical response of~\eqref{eq:bp_surface_phase}. The phase MAE between the two curves of $0.14^\circ$ matches the spatial error $\varepsilon_s$ established in Section~\ref{subsec:res_semi_infinite_solid}, thus confirming that the spatial discretization criterion transfers unchanged to the bounded geometry. The solution also reproduces the limiting values established in Table~\ref{tab:phasevalues}: thermally thin limit approaching $-90^\circ$, thermally thick limit approaching $-45^\circ$, and blind frequency at $L/\mu=\pi/2$ where the surface phase crosses $-45^\circ$.

\subsection{Subsurface defect}
\label{subsec:res_blind_freq}
The final benchmark applies the protocol to subsurface defects. As no closed-form solution exists for a defect of finite lateral extent (Section~\ref{subsec:blind_freq}), the aim of this section is to demonstrate that the understandings from the previous sections extend to a three-dimensional geometry, and that the recovered blind-frequency relation depends on respecting the discretization criterion. Furthermore, the material is changed from AlMg3 to Al to show that the established discretization guideline is material-agnostic. The geometry consist of a flat-bottom hole defects (Fig.\ref{fig:benchmarks}(c)) of radius $r=$ \qty{0.5}{\milli\meter} at varying depths $d$ of \qtyrange{0.2}{1.4}{\milli\meter} in a sample with thickness $L$ of \qty{6}{\milli\meter}. Taking advantage of defect symmetry, harmonic simulations were carried out on an axisymmetric model with a sample radius of \qty{2}{\milli\meter}. Two meshing strategies were compared: the $\mu$-scaled protocol, in which the element size is rescaled with frequency to hold $a=17$ elements per diffusion length, and a constant mesh of fixed element size (\qty{80}{\micro\meter} in the defect region and \qty{0.15}{\milli\meter} in the sound region), representing a constant mesh across the frequency sweep. For each defect, the blind frequency $f_b$ was identified as the zero crossing of the phase contrast between the defect and the sound regions, located by linear interpolation of the sweep contrast. 

\begin{figure}
	\centering
	\includegraphics[width=\columnwidth]{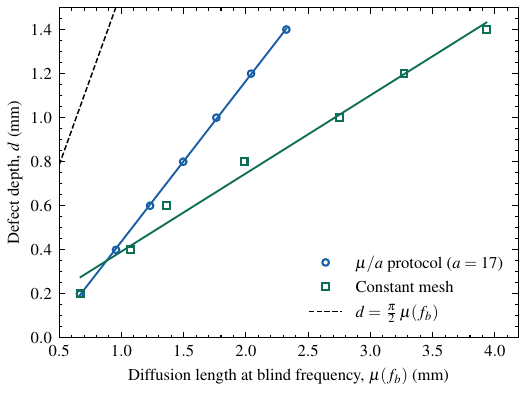}
	\caption{Recovered defect depth ($d$) versus diffusion length at the blind frequency ($\mu(f_b)$), for the $\mu$-scaled protocol ($a=17$) and a constant mesh, compared with the case when the phase contrast reference is thermally thick~\eqref{eq:blind_freq}. The protocol recovers a stable relation  $d=0.73\mu(f_b)-0.29~\text{mm}$; constant mesh yields a non-linear response}
	\label{fig:defect_blind_freq}
\end{figure}

Fig.\ref{fig:defect_blind_freq} shows the obtained defect depth $d$ against the diffusion length at the blind frequency $\mu(f_b)$ for both the meshing strategies. The $\mu$-scaled protocol yields a well-defined linear relation of $d=0.73\mu(f_b)-0.29~\text{mm}$, whereas the constant mesh yields a non-linear response from otherwise identical simulations. Both relations lie well below the expected relation of $d=(\pi/2)\mu(f_b)$. This departure from the expected result can be explained by the fact that the ratio of the defect radius to the diffusion length, $r/\mu<1$, across the defect set, so lateral conduction around the defect contributes throughout \cite{maQuantitativeDepthEstimation2025} and the finite thickness sound region provides a non-ideal reference (Section~\ref{subsec:blind_freq}). The non-zero intercept value of the protocol fit confirms that the obtained relationship is an operational approximation rather than an exact identity. The central result of this study is that the obtained relation, and hence any depth inferred from it, depends on fulfilling the discretization criterion. A compliant discretization produces a stable and repeatable relation.
\subsection{Recommended Discretization Protocol}
The preceding benchmark results and the corresponding discussions yield a discretization protocol for FE simulations of thermal waves. The protocol is expressed in dimensionless mesh $a = \Delta x/\mu$ and time-step measures $N_t = T/\Delta_t$, where $T=1/f$ so it applies independently of materials and excitation frequencies.

\noindent\textit{Analysis type}:
The choice between harmonic and transient analysis follows from the quantity of interest. When only the periodic steady-state response (phase and amplitude values) is required, harmonic analysis is both faster and free of temporal discretization error. Transient analysis is required only when the approach to periodic steady state is of interest-the initial response to the mean component of the excitation source, the number of excitation periods needed to reach the periodic steady state, or thermal losses during heat-up. Even then, the transient cost can be reduced by first applying the DC component alone to establish the steady temperature, before superimposing the modulated transient.

\noindent\textit{Spatial resolution}:
The surface phase error follows $|\varepsilon_s|\approx 12.55^\circ\cdot a^{-1.95}$ (Fig.\ref{fig:spatial_conv}). $a\ge10$ elements per diffusion length keeps the surface error below $0.15^\circ$. As amplitude error converges earlier at a coarse mesh than phase error (Section~\ref{subsec:res_semi_infinite_solid}), the mesh quality must be judged on phase error to avoid false convergence.

\noindent\textit{Temporal resolution}:
At a fixed mesh, the temporal phase error follows $|\varepsilon_t|=90^\circ/N_t$ (Fig.~\ref{fig:temporal_conv}), independent of the mesh. Thus, for a given phase tolerance of $\varepsilon$, the sampling requirement is $N_t\ge90^\circ/\varepsilon$. 

\noindent\textit{Refinement strategy}:
The spatial and temporal errors are separable in the periodic steady state and must be controlled independently. For transient analysis, refining the time step drives the solution toward the harmonic solution on the same mesh, retaining the spatial error associated with the mesh size. For harmonic analysis, convergence to the analytical solution is achieved by refining the mesh, whereas for transient solutions, convergence is achieved by refining both the mesh and solution time steps. The associated errors are discussed in the previous two points. 

\noindent\textit{Graded meshing}:
The uniform mapped meshes used here resolve the diffusion length everywhere in the domain, which is unnecessary because the thermal-wave amplitude decreases exponentially along the diffusion length. A graded mapped mesh, refined to satisfy $a\ge 10$ within one to two diffusion lengths of the region of interest, followed by a gentle geometric growth in the element size, retains the required resolution at reduced element count. 

\noindent\textit{Scope and transferability}:
The procedure of benchmarking against the closed-form solution, measuring the convergence order, and setting $a$ and $N_t$ from the desired tolerance transfers to any solver and geometry as demonstrated here across different cases. The numerical constants, however, are specific to the current configuration of linear thermally conducting brick elements and a backward-Euler integrator. For a different element type or time-integration scheme, remeasure these constants using the same procedure; the dimensionless framework and the reference-selection rules (Fig.\ref{fig:temporal_conv}) remain unchanged.


\section{Conclusions}
\label{sec:conclusions}
This work provided an analytically benchmarked quantification of amplitude and phase errors in FE simulations of thermal waves. Across three benchmark problems of increasing complexity, the spatial and temporal error contributions were isolated and, wherever a closed-form solution exists, the obtained results were verified against it. 

Refining either the mesh or the time-step alone increases the total error; each error axis must therefore be benchmarked against a reference that isolates it. Based on the analysis of the results, a spatial and temporal discretization protocol is suggested that addresses the key points such as selection of the correct analysis type for a given quantity of interest, the expected resolution for a selected spatial and temporal discretization, reducing the element count with a graded mesh, and, importantly, the scope and transferability of the recommended protocol. 

With discretization errors now bounded against a closed-form reference, the proposed framework can be applied to geometries with no analytical solution and subsequently validated against experiments, thus making the benchmarked simulations a quantitative measurement tool. 

\section*{Acknowledgment}
Part of the presented work has been funded by SmartMan project from the Federal Ministry for Economic Affairs and Climate Action (Grant No. 13IK033E). The authors appreciate this funding.

\bibliographystyle{IEEEtran} 
\bibliography{references}
\end{document}